\documentclass[pre,aps,twocolumn,showpacs,superscriptaddress,floatfix]{revtex4-2}

\usepackage{mathtools}
\usepackage[usenames,dvipsnames]{xcolor}
\usepackage{amsmath}
\usepackage{amssymb,mathrsfs}
\usepackage{graphicx}
\usepackage[colorlinks=true]{hyperref}
\usepackage{soul}
\usepackage{xcolor}

\newcommand{\prt}{\partial}
\newcommand{\al}{\alpha}

\newcommand{\om}{\omega}

\newcommand{\ox}{\overline{x}}

\newcommand{\occ}{\overline{c}}

\begin{document}

\title{Number of solitons produced from a large initial pulse in the generalized
NLS dispersive hydrodynamics theory}

\author{L. F. Calazans de Brito}  
  \affiliation{Higher School of Economics, Moscow, Russia}

\author{A. M. Kamchatnov}  \affiliation{Higher School of Economics, Moscow, Russia}
 \affiliation{Institute of Spectroscopy,
  Russian Academy of Sciences, Troitsk, Moscow, 108840, Russia}

\begin{abstract}
We show that the number of solitons produced from an arbitrary initial pulse
of the simple wave type can be calculated analytically if its evolution is
governed by a generalized nonlinear Schr\"{o}dinger equation provided this
number is large enough. The final result generalizes the asymptotic formula
derived for completely integrable nonlinear wave equations like the
standard NLS equation with the use of the inverse scattering transform method.
\end{abstract}

\pacs{05.45.Yv, 42.65.Tg, 47.35.Fg}


\maketitle

\section{Introduction}

It is well known that if a nonlinear wave system supports propagation of solitons
with some polarity (bright or dark ones), then a large enough initial pulse with the
same polarity evolves at asymptotically large times to a sequence of $N$ separate
solitons with some amount of linear radiation. The number $N$ of soliton depends on the
initial data and for $N\gg1$ the energy contained in linear radiation is negligibly
small compared with the solitons' energy in the main approximation with respect to
the small parameter $1/N\ll1$. Consequently, the number of solitons $N$ is one of the
most important characteristics of the initial pulse. Besides that, it is relatively
easy to measure $N$ experimentally. Thus, the possibility to predict the number of solitons 
from the initial data is an important task in the theory of nonlinear waves.

If dynamics of the system under consideration is described by a completely integrable wave
equation, then the number of solitons is equal to the number of eigenvalues of the associated
linear spectral problem (see, e.g., \cite{nmpz-84,newell-85}). Consequently, for large
number $N\gg1$ one can apply the quasiclassical  WKB method to the linear spectral problem
and obtain the asymptotic formula for $N$. For example, in case of the famous KdV equation
\begin{equation}\label{kdv}
  u_t+6uu_x+u_{xxx}=0
\end{equation}
such a formula was first obtained by Karpman \cite{karpman} and it reads
\begin{equation}\label{karpman-67}
  N\approx\frac1{\pi}\int_{-\infty}^{\infty}\sqrt{u_0(x)}\,dx, \qquad N\gg1,
\end{equation}
where $u_0(x)$ is the initial distribution of the amplitude $u$ and $N$ is the number of
eigenvalues of the Schr\"{o}dinger spectral problem \cite{ggkm-67,nmpz-84,newell-85}
with the ``potential'' $u_0(x)$. A similar formula was derived in Ref.~\cite{JLML-99}
for the NLS equation
\begin{equation}\label{NLS}
  i\psi_t+\frac12\psi_{xx}-|\psi|^{2}\psi=0
\end{equation}
associated with the Zakharov-Shabat spectral problem \cite{zs-73}. This formula is 
conveniently formulated
in terms of the initial data for the hydrodynamics-like form of Eq.~(\ref{NLS}) obtained by
means of the Madelung substitution
\begin{equation}\label{madelung}
  \psi(x,t)=\sqrt{\rho(x,t)}\exp(i\phi(x,t)),\qquad \phi_x=u(x,t),
\end{equation}
so that the NLS equation reduces to the system
\begin{equation}\label{nls-system}
  \begin{split}
  & \rho_t+(\rho u)_x=0,\\
  & u_t+uu_x+\rho_x+\left(\frac{\rho_x^2}{8\rho^2}-\frac{\rho_{xx}}{4\rho}\right)_x=0.
  \end{split}
\end{equation}
In the context of the Gross-Pitaevskii theory \cite{gross-61,pit-61} of Bose-Einstein
condensates of diluted gases, $\rho$ has the meaning of the gas density and $u$ is
its flow velocity. This interpretation becomes especially clear in the limit of large
pulses with their characteristic size $l\gg1$. Then the space derivative has the order
of magnitude $\prt_x\sim l^{-1}\ll1$ and the last term in the second equation (\ref{nls-system})
can be neglected. Hence we arrive at equations of Euler hydrodynamics
\begin{equation}\label{Euler}
  \rho_t+(\rho u)_x=0,\qquad u_t+uu_x+\rho_x=0
\end{equation}
for a compressible fluid with the equation of state $P=\frac12\rho^2$,
where $P$ plays the role of ``pressure''.

Dynamics of such a gas is conveniently described with the use of the variables
\begin{equation}\label{rim_inv}
  r^{\pm}=\frac12u\pm\sqrt{\rho}
\end{equation}
called Riemann invariants, so that Eqs.~(\ref{Euler}) acquire a diagonal form
\begin{equation}\label{diagonal}
\begin{split}
  & \frac{\prt r^+}{\prt t}+\frac12(3r^++r^-)\frac{\prt r^+}{\prt x}=0,\\
  & \frac{\prt r^-}{\prt t}+\frac12(r^++3r^-)\frac{\prt r^-}{\prt x}=0.
  \end{split}
\end{equation}
The Gross-Pitaevskii equation has dark soliton solutions \cite{tsuzuki} propagating
in the form of dips along a uniform background $\rho=\overline{\rho},u=\overline{u}$.
If the initial distributions $\overline{\rho}_0(x),\overline{u}_0(x)$ correspond to a
large dip in the density, $0<\overline{\rho}_0(x)<\overline{\rho}$ and
$\overline{u}_0(x)\to0$ as $|x|\to\infty$, then we can transform them to distributions
of the Riemann invariants $r_0^{\pm}(x)=\overline{u}_0(x)/2\pm\sqrt{\overline{\rho}_0(x)}$.
The asymptotic formulas for a number of dark solitons moving in the positive or negative
direction are given, correspondingly, by the equations \cite{JLML-99}
\begin{equation}\label{NLS-number}
   N_{\pm}=\frac1{\pi}\int_{-\infty}^{\infty}\sqrt{(\pm\overline{r}-r_0^+(x))
  (\pm\overline{r}-r_0^-(x))}\,dx,
\end{equation}
where $\overline{r} =\sqrt{\overline{\rho}}$.

Less rigorous simple approach to this problem was suggested in Ref.~\cite{kkb-01} for the
whole AKNS hierarchy \cite{akns} associated with $2\times2$-matrix spectral problem
written in an equivalent scalar form \cite{kk-02}. Naturally, in case of Eq.~(\ref{NLS})
it reproduces Eq.~(\ref{NLS-number}) (see Ref.~\cite{kkb-02}). The approach of
Refs.~\cite{kkb-01,kkb-02} is based on the supposition that the solution of the linear
spectral problem (the Baker-Akhiezer function) corresponding to the periodic nonlinear
wave keeps its form for slightly modulated waves and can be formally continued to the
initial state where it becomes a quasiclassical eigenfunction of the spectral problem.

For not completely integrable equations the associated linear spectral problem does not
exist and the above approach becomes impossible. Since the process of formation of
solitons from an initially smooth pulse is universal from physical point of view and,
generally speaking, it does not depend on the fact whether the wave equation is
completely integrable or not, other ideas are necessary. An alternative approach was
suggested in Refs.~\cite{egkkk-07,egs-08} and it was based on the Gurevich-Pitaevskii
theory  of dispersive shock waves (DSWs). According to this theory, transformation of an
initially smooth pulse goes through several stages. In the first stage, the pulse changes
its form remaining smooth up to the moment of wave breaking when the dispersionless
approximation breaks down due to appearance of infinite space derivatives (``gradient
catastrophe''). After the wave breaking moment, the second stage starts during which
the pulse in the simplest case consists of two parts---a smooth part for which the
dispersionless approximation remains correct, and the DSW part represented by a modulated
periodic solution of the nonlinear wave equation under consideration where the modulation
parameters obey the Whitham modulation equations \cite{whitham-65} (see also review
articles \cite{eh-16,kamch-21c} for more details). In Whitham approximation,
the boundary between the two parts is sharp and it is called the small-amplitude edge of
the DSW. At last, in the third asymptotic stage of the pulse's evolution its smooth part
becomes negligibly small and the whole pulse evolves to a sequence of separate solitons.
The method of Refs.~\cite{egkkk-07,egs-08} was based on the assumption that the solution
of the Whitham equations at the small-amplitude edge of the DSW can be formally prolonged
to the smooth part of the pulse. The necessary parameters of the DSW at its small-amplitude
edge can be obtained in the important case of initial simple waves with one of the Riemann
invariants constant by El's method \cite{el-05}. Although the results of this approach
were confirmed in several particular cases by their comparison with the results of numerical
solutions, it hardly can be considered as obvious, because the Whitham equations are
obtained by averaging over fast nonlinear wave oscillations, whereas there are no any 
oscillations in the smooth part of the pulse and introduction of their wave vector here
looks quite artificial. Therefore other approach based on clearer physical ideas seems
very desirable.

Recently a new approach has been suggested in Refs.~\cite{kamch-21c,kamch-20a}. It is based
on an old remark of Gurevich and Pitaevskii \cite{gp-87} that at the mentioned above second
stage of evolution the number $N(t)$ of oscillations (wave crests) inside the DSW part
increases with time according to the equation
\begin{equation}\label{gp-N}
  \frac{dN}{dt}=\frac1{2\pi}k(v_g-v_{ph}),
\end{equation}
where $k$ is the wave number and $v_g,v_{ph}$ are the group and the phase velocities,
correspondingly, of the wave at the small-amplitude edge of the DSW. If we denote the
background amplitude at this edge as $u$ then the dependence $k=k(u)$ can be found by
El's method and the dependence $t=t(u)$ along the path of the small-amplitude edge can
be found by the method of Ref.~\cite{kamch-19a}. As a result, Eq.~(\ref{gp-N}) can be
integrated along the path of the small-amplitude edge and in the limit $t\to\infty$
we obtain the total number of crests in the DSW which evolve eventually to separate
solitons. As was shown in Refs.~\cite{kamch-20a,kamch-21a} for several simple particular
cases, this method yields the formulas for the number of solitons coinciding with those 
derived by the method of Refs.~\cite{egkkk-07,egs-08}, so in these cases the formulas 
are justified by a more reliable approach.

In this paper, we consider the problem of calculation of the number of solitons
produced from a simple-wave type of an initial pulse which evolves according to the
generalized NLS (gNLS) equation
\begin{equation}\label{gNLS}
  i\psi_t+\frac12\psi_{xx}-\frac1p|\psi|^{2p}\psi=0.
\end{equation}
It reduces to the standard NLS equation (\ref{NLS}) for $p=1$, but for $p\neq1$ it
is not completely integrable. Various forms of nonlinearity in the gNLS equation
are used in nonlinear physics. In particular, the regime $2/3 < p < 1$ describes
the superfluid BEC–Bardeen–Cooper–Schrieffer transition in ultracold Fermi gases
\cite{kz-08} and the case $p=2/3$ of the so-called unitary limit has drawn considerable
attention (see, e.g., \cite{kkt-20} and reference therein). Besides that, Eq.~(\ref{gNLS})
with different values of $p$ are very useful and convenient for the study of various
properties of dispersive Euler hydrodynamics \cite{hoefer-14} since it models both
integrable and non-integrable situations amenable to analytical study.
Here we show that the formula for the number of solitons can be derived by the direct
method of Refs.~\cite{kamch-21c,kamch-20a} and this formula can be used in
applications and general nonlinear physics investigations.

\section{Basic formulas}

We shall present here the basic formulas necessary for derivation of the expression
for the number $N$ of solitons.

\subsection{Simple waves}

We assume that at the dispersionless stage of evolution before the wave breaking moment
the pulse has a form of a simple wave. Actually, this is not a serious restriction
imposed on the form of the pulse, since for quite arbitrary initial conditions the
dispersionless evolution leads in a natural way to splitting the pulse to two pulses
propagating in opposite directions and they both are of the simple wave type. 
Besides that, simple wave pulses can be
created by a proper choice of the initial conditions as it happens, for example,
in the flow caused by a unidirectional motion of a piston (see, e.g., \cite{LL-6}).

Equations of the dispersionless approximation for Eq.~(\ref{gNLS}) can be obtained by
means of substitution (\ref{madelung}) into it, separation of real and imaginary parts
and neglecting the dispersive terms with higher order space derivatives. As a result,
we obtain the Euler equations
\begin{equation}\label{Euler-2}
  \rho_t+(\rho u)_x=0,\qquad u_t+uu_x+\rho^{p-1}\rho_x=0
\end{equation}
corresponding to the equation of state $P=\rho^{p+1}/(p+1)$. It is convenient to define
the ``sound velocity'' variable
\begin{equation}\label{eq2}
  c=\sqrt{dP/d\rho}=\rho^{p/2}.
\end{equation}
Then for the Riemann invariants
\begin{equation}\label{riemann-2}
  r_{\pm}=\frac{u}{2}\pm\frac{c}{p}
\end{equation}
Eqs.~(\ref{Euler-2}) transform to
\begin{equation}\label{eq15-d}
  \prt_t r_{\pm}+(u\pm c)\prt_x r_{\pm}=0,
\end{equation}
where the characteristic velocities $v_{\pm}=u\pm c$ have simple physical meaning---they
correspond to waves propagating with the sound velocity $c=c(\rho)$ downstream or
upstream, correspondingly.

\begin{figure}[t]
\begin{center}
\includegraphics[width=8.5cm]{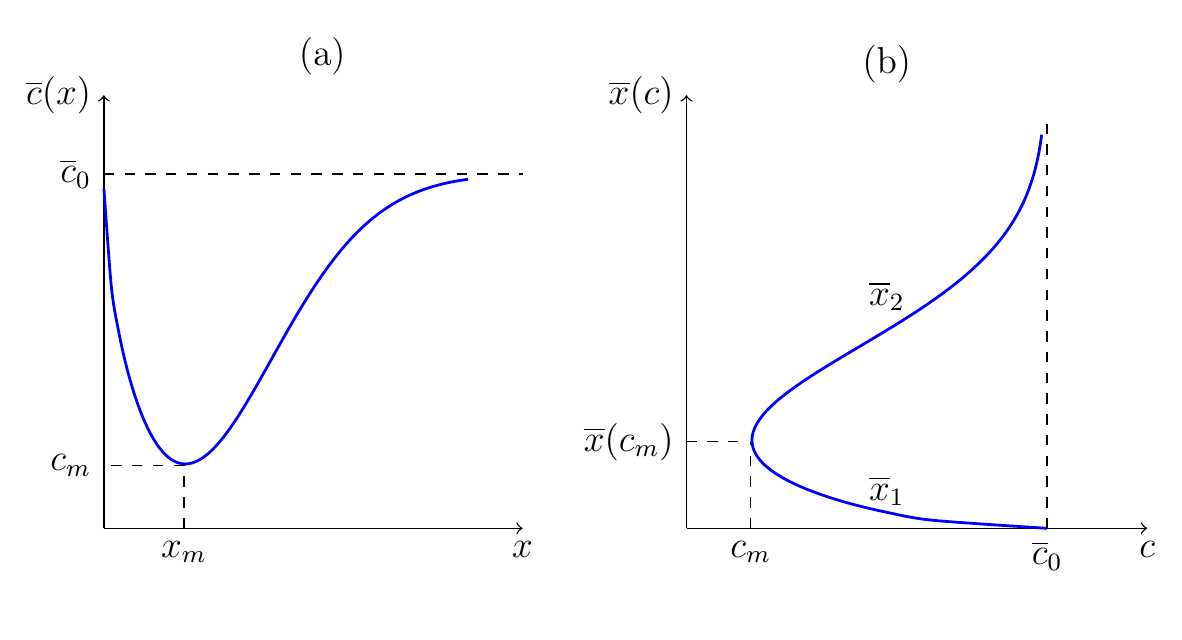}
\caption{
(a) The initial distribution of the local sound velocity $\overline{c}(x)$.
(b) The inverse function $\ox(c)$ has two branches $x_1(c)$ and $x_2(c)$.
}
\label{fig1}
\end{center}
\end{figure}

In a simple wave one of the Riemann invariants is constant and, to be definite,
we assume that $r_-=u/2-c/p=-\occ_0/p$, where $\occ_0$ is the sound velocity far from
the localized initial pulse. Then the flow velocity $u(x,t)$ is expressed  in terms
of the distribution $c(x,t)$ of the local sound velocity,
\begin{equation}\label{eq16-d}
  u(x,t)=\frac2{p}(c(x,t)-\occ_0),
\end{equation}
hence $r_+=(2c(x,t)-\occ_0)/p$ and Eq.~(\ref{eq15-d}) for $r_+$ is transformed to
the equation
\begin{equation}\label{eq17-d}
  \frac{\prt c}{\prt t}+\frac1p[(2+p)c-2\occ_0]\frac{\prt c}{\prt x}=0.
\end{equation}
This is the well known Hopf equation (see, e.g., \cite{whitham}) and its solution
reads
\begin{equation}\label{eq18-d}
  x-\frac1p[(2+p)c-2\occ_0]t=\ox(c),
\end{equation}
where $\ox(c)$ is the function inverse to the initial distribution $c=\occ(x)$
of the sound velocity (\ref{eq2}). We are interested in the initial distributions
in the form of localized dips in the uniform state with $c=\occ_0$ (see Fig.~\ref{fig1}(a)).
Hence, the inverse function has two branches $x=\ox_{1,2}(c)$ (see Fig.~\ref{fig1}(b)), and
each branch gives the solution (\ref{eq18-d}) which defines in implicit form the
function $c=c(x,t)$. Evolution of the profile $c=c(x,t)$ leads to wave breaking and,
to simplify the notation, we assume that the wave breaking moment corresponds to $t=0$.

\subsection{Dispersion relation and El's equation}

After the wave breaking moment a DSW is generated and its small amplitude edge starts
its motion along the smooth part of the pulse. Actually, this edge is represented by
a linear wave packet propagating with the group velocity of linear waves. The
corresponding dispersion relation can be easily found after linearization of Eq.~(\ref{gNLS})
and it can be written in the form (see, e.g., \cite{hoefer-14})
\begin{equation}\label{eq7}
  \om(k,c)=k\left(u\pm\sqrt{c^2+k^2/4}\right). 
\end{equation}
The background distribution $c=c(x,t)$ is changing according to the Hopf equation
(\ref{eq17-d}) which should be combined with the Hamilton equations
\begin{equation}\label{eq20-d}
  \frac{dx}{dt}=\frac{\prt\om}{\prt k},\qquad \frac{dk}{dt}=-\frac{\prt\om}{\prt x}
\end{equation}
for the packet's motion. As a simple consequence of these three equations we easily obtain
\cite{kamch-20a} the equation
\begin{equation}\label{eq21-d}
  \frac{dk}{dc}=\frac{(\prt\om/\prt c)_k}{u+c-(\prt\om/\prt k)_c}.
\end{equation}
This equation was first obtained by El \cite{el-05} from the small-amplitude limit of
the Whitham equations. Its derivation from Hamilton's equations demonstrates its much
wider applicability region. For example, it was applied in Ref.~\cite{ks-21} to
propagation of localized wave packets along arbitrary large scale smooth simple waves.
In case of motion of the small-amplitude edge of a DSW it should be solved with the
initial condition
\begin{equation}\label{init-cond}
  k(\occ_0)=0
\end{equation}
which means that the wave breaking occurs at the boundary with the undisturbed region where
$c=\occ_0$ and in the Whitham approximation the DSW at the initial wave breaking
moment shrinks to a point without any oscillations inside.

To solve this equation, it is convenient to introduce the variable \cite{egkkk-07}
\begin{equation}\label{eq8}
  \al(c)=\sqrt{1+\frac{k^2(c)}{4c^2}},\qquad k(c)=2c\sqrt{\al^2(c)-1},
\end{equation}
so that Eq.~(\ref{eq21-d}) transforms to
\begin{equation}\label{eq9}
  \frac{d\al}{dc}=-\frac{(1+\al)(2/p-1+2\al)}{c(1+2\al)},
\end{equation}
which can be easily solved with the initial condition $\al(\occ_0)=1$
to give \cite{hoefer-14}
\begin{equation}\label{eq10}
  \frac{c(\al)}{\occ_0}=\left(\frac2{1+\al}\right)^{\frac{p}{3p-2}}
  \left(\frac{2+p}{2-p+2p\al}\right)^{\frac{2(p-1)}{3p-2}}.
\end{equation}
The function $c=c(\al)$ can be inverted in three particular cases \cite{hoefer-14}:
\begin{equation}\label{eq11}
  \begin{split}
  & p=1,\quad \al(c)=\frac{2\occ_0}c-1,\\
  & p=2,\quad \al(c)=\frac12\left(\sqrt{1+8\left(\frac{\occ_0}c\right)^2}-1\right),\\
  & p=\frac12,\quad \al(c)=\frac1{16}\left(5\sqrt{\frac{\occ_0}c\left(25\frac{\occ_0}c-16\right)}
  +25\frac{\occ_0}c-24\right).
  \end{split}
\end{equation}
Substitution of these formulas into Eq.~(\ref{eq8}) yields explicit expressions for
the function $k(c)$.

\subsection{Path of the small amplitude edge}

The small amplitude edge propagates with the group velocity
\begin{equation}\label{eq26-d}
  \frac{dx}{dt}=\frac{d\om}{dk}=\frac2p(c-\occ_0)+c(\al(c)-\al^{-1}(c)).
\end{equation}
According to Ref.~\cite{kamch-19a}, this equation must be compatible with the solution
(\ref{eq18-d}), where $c$ equals to the local value of the sound velocity at the
point of location of the wave packet. This condition leads to the linear equation
\begin{equation}\label{eq27}
  \frac{c(\al-1)(2\al+1)}{\al}\frac{dt}{dc}-\frac{2+p}p t=\ox'(c),
\end{equation}
which should be solved first with the initial condition $t(\occ_0)=0$. Easy
integration yields
\begin{equation}\label{eq28}
\begin{split}
  t=t_1(c)&=\frac{1}{\sqrt{c(\al(c)-1)}}\left(\frac{\al(c)+1}{2-p+2p\al(c)}\right)^{\frac{1}{3p-2}}\\
  &\times \int_{\occ_0}^cdc'\frac{\ox_1(c')}{\sqrt{c'(\al(c')-1)}(2\al(c')+1)}\\
  &\times \left(\frac{2-p+2p\al(c')}{\al(c')+1}\right)^{\frac{1}{3p-2}}
  \end{split}
\end{equation}
for the motion of the wave packet along the branch $x_1(c)$: at the moment $t$ the small-amplitude
edge is located at the point with the local value of the sound velocity $c$. The expression
(\ref{eq28}) is correct up to the moment $t_m=t_1(c_m)$ when the edge reaches  the point with
the minimal value $c_m$ of the local sound velocity. It is worth noticing that the dispersionless
solution (\ref{eq18-d}) preserves the minimal value $c_m$ in the distribution $c=c(x,t)$, so $c_m$
does not depend on $t$. For $t>t_m$ the edge propagates along the solution (\ref{eq18-d})
corresponding to the second branch $\ox_2(c)$. Accordingly, Eq.~(\ref{eq27}) should be solved
with the initial condition $t(c_m)=t_m$ and the solution reads
\begin{equation}\label{eq29}
\begin{split}
  t=t_2(c)&=\frac{1}{\sqrt{c(\al(c)-1)}}\left(\frac{\al(c)+1}{2-p+2p\al(c)}\right)^{\frac{1}{3p-2}}\\
  &\times \Bigg\{\int_{\occ_0}^{c_m}dc'\frac{\ox_1(c')}{\sqrt{c'(\al(c')-1)}(2\al(c')+1)}\\
  &\times \left(\frac{2-p+2p\al(c')}{\al(c')+1}\right)^{\frac{1}{3p-2}}\\
  & +\int_{c_m}^{c}dc'\frac{\ox_2(c')}{\sqrt{c'(\al(c')-1)}(2\al(c')+1)}\\
  &\times \left(\frac{2-p+2p\al(c')}{\al(c')+1}\right)^{\frac{1}{3p-2}}
  \Bigg\}.
  \end{split}
\end{equation}
Substitution of the above two formulas into $x(c)=\ox_{1,2}(c)+\frac1p[(2+p)c-2\occ_0]t_{1,2}(c)$
gives the coordinate of the edge at the moment $t(c)$, but we do not need these expressions here.

\section{Number of solitons}

Integration of formula (\ref{gp-N}) with the use of the dispersion relation (\ref{eq7})
gives the total number of wave crests containing in the DSW,
\begin{equation}\label{eq30}
  N=\frac1{8\pi}\int_0^{\infty}\frac{k^3}{\sqrt{c^2+k^2/4}}\,dt,
\end{equation}
so our task is to evaluate this integral. To this end we pass to integration over $c$ and
substitute (\ref{eq8}) for $k=k(c)$,
\begin{equation}\label{eq31}
  N=\frac1{\pi}\int\frac{c^2}{\al(c)}(\al^2(c)-1)^{3/2}\frac{dt}{dc}dc.
\end{equation}
In fact, this integral consists of two parts: first with integration from $\occ_0$ to $c_m$
with $t=t_1(c)$ and second with integration from $c_m$ to $\occ_0$ with $t=t_2(c)$.
The derivative $dt/dc$ can be excluded with the use of Eq.~(\ref{eq27}) and as a result
we obtain
\begin{equation}\label{eq32}
\begin{split}
  N=&\frac1{\pi}\int_{\occ_0}^{c_m}\frac{c(\al-1)^{1/2}(\al+1)^{3/2}}{2\al+1}\\
  &\times \left[\frac{2+p}p(t_2-t_1)+\ox_2-\ox_1\right]dc.
  \end{split}
\end{equation}
Integration over $c$ can be replaced by integration over $\al$ with help of Eq.~(\ref{eq9}),
\begin{equation}\label{eq33}
\begin{split}
  N=&\frac{p}{\pi}\int_{1}^{\al_m}\frac{c^2(\al)(\al^2-1)^{1/2}}{2-p+2p\al}\\
  &\times \left[\frac{2+p}p(t_2-t_1)+\Phi_2-\Phi_1\right]d\al,
  \end{split}
\end{equation}
where $\al_m=\al(c_m)$ and $\Phi(\al)\equiv\left.\ox'(c)\right|_{c=c(\al)}$.

When we substitute here the expressions (\ref{eq28}) and (\ref{eq29}) for $t_1$ and $t_2$
as well as Eq.~(\ref{eq10}) for $c(\al)$, we obtain the double integral which can be
simplified by integration by parts:
\begin{equation}\nonumber
  \begin{split}
  &\frac{2+p}{\pi}\int_{1}^{\al_m}\frac{c^2(\al)(\al^2-1)^{1/2}}{2-p+2p\al}(t_2-t_1)d\al\\
  &=\frac1{\pi}\int_{c_m}^{\occ_0}\frac{c\al\sqrt{\al^2-1}(\ox'_2-\ox'_1)}{2\al+1}dc.
  \end{split}
\end{equation}
The remaining terms are transformed to
\begin{equation}\nonumber
  \begin{split}
  &\frac{p}{\pi}\int_{1}^{\al_m}\frac{c^2(\al)(\al^2-1)^{1/2}}{2-p+2p\al}(\Phi_2-\Phi_1)d\al\\
  &=\frac1{\pi}\int_{c_m}^{\occ_0}\frac{c(\al+1)\sqrt{\al^2-1}(\ox'_2-\ox'_1)}{2\al+1}dc,
  \end{split}
\end{equation}
and the sum of the above two expressions yields
\begin{equation}\label{eq34}
  N=\frac1{\pi}\int_{c_m}^{\occ_0}c(\al^2-1)^{1/2}(\ox'_2-\ox'_1)dc.
\end{equation}
At last, we replace integration over the interval $c_m\leq c\leq\occ_0$ by integration
over $x$ and take into account Eq.~(\ref{eq8}) to obtain the final formula
\begin{equation}\label{N-solitons}
  N=\frac1{2\pi}\int_{-\infty}^{\infty}k(\occ(x))dx,
\end{equation}
where $\occ(x)$ is the initial distribution of the local sound velocity. In this
expression the function $k(c)$ is obtained by substitution of the function $\al(c)$
into Eq.~(\ref{eq8}). For example, in case $p=1$ of the standard NLS equation we
have for $\al(c)$ the first expression in Eq.~(\ref{eq29}), hence
$k(c)=4\sqrt{\occ_0(\occ_0-c)}$ and Eq.~(\ref{N-solitons}) reduces to
\begin{equation}\label{eq36}
  N=\frac2{\pi}\int_{-\infty}^{\infty}\sqrt{\occ_0(\occ_0-\overline{c}(x))}dx.
\end{equation}
This formula coincides with formula (\ref{NLS-number}) for $N_+$, since for the
simple wave initial conditions we have $r_-=-\occ_0,r_+=2\occ(x)-\occ_0$.

We have obtained Eq.~(\ref{N-solitons}) by direct calculation without artificial
introduction of the distribution of wave numbers $k(\occ(x))$ along a smooth
initial state, as it was done in Refs.~\cite{egkkk-07,egs-08}. Thus, our derivation
makes the assumption about existence of such a distribution quite plausible
(see \cite{kamch-20a}).

\section{Comparison with numerical solutions}

\begin{figure}[t]
\begin{center}
\includegraphics[width=8cm]{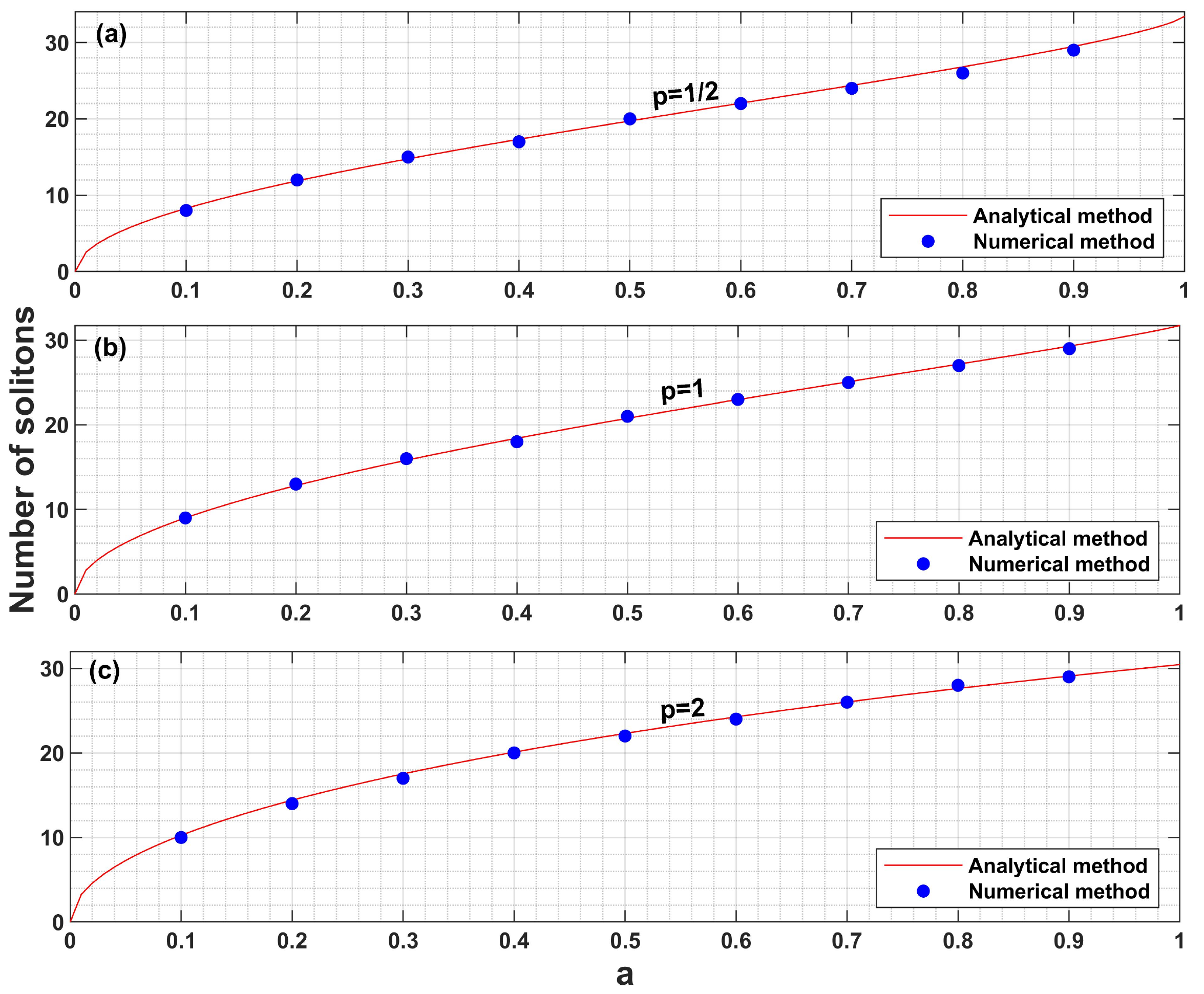}
\caption{
The number of solitons produced from the initial pulse (\ref{eq37}) for $l=20$ and different values of $a$. Solid lines correspond to
(a) $p=1/2$; (b) $p=1$; and (c)  $p=2$ calculated according
to the analytical formula (\ref{N-solitons}).
}
\label{fig2}
\end{center}
\end{figure}

Here we support our analytical theory by comparison of its predictions with the results
of numerical solutions of the gNLS equation. We take the initial distribution of
the local sound velocity $c=\rho^{p/2}$ in the form
\begin{equation}\label{eq37}
  \overline{c}(x)=\left[1-\frac{a}{\cosh^2(x/l)}\right]^{p/2},
\end{equation}
where $a$ denotes the depth of the initial dip in the density distribution and
$l$ is its half-length. The
initial disturbance must be a simple wave in our approach, so we define the
initial distribution of the flow velocity by the formula
\begin{equation}\label{eq38}
  \overline{u}_0(x)=\frac2p(\overline{c}(x)-\overline{c}_0).
\end{equation}
Equations (\ref{madelung}) allow us to find the initial field $\psi(x,0)$.

In our numerical experiments we choose $l=20$ and change $a$ in the interval
$0.1\leq a\leq 0.9$ with the step $\Delta a=0.1$. Such a choice makes the 
numerical calculations not too time consuming and clearly demonstrates the
dependence of the number of solitons on $a$. We have done these
calculations for $p=1/2,1$ and 2, when the function $k(\overline{c}(x))$ is
given by the explicit formulas (\ref{eq8}) and (\ref{eq11}). The results of our
calculations are presented in Fig.~\ref{fig2}. As one can see, the agreement is
very good.

\section{Conclusion}

We have shown that the method of calculation of the number of solitons produced
from an initial pulse of the simple wave type works very well for the
generalized NLS equation having various physical application. The resulting
formula (\ref{N-solitons}) has the structure suggested earlier in
Ref.~\cite{egkkk-07,egs-08} on the basis of some suppositions about the properties
of solutions of Whitham modulation equations and our derivation makes these
suppositions quite plausible. Thus, the developed method and the general
formula (\ref{N-solitons}) become a useful tool for predictions of the number
of solitons in experiments performed with media whose evolution is described by
nonlinear wave equations not belonging to a specific class of completely
integrable equations.

\begin{acknowledgments}
The work was supported by the Foundation for the Advancement of Theoretical
Physics and Mathematics ``BASIS''.
\end{acknowledgments}

\end{document}